\documentclass[aps, pra, twocolumn,showpacs,preprintnumbers,amsmath,amssymb]{revtex4}

\usepackage{graphicx}
\usepackage{dcolumn}
\usepackage{bm}

\newcommand{\bR}{\mbox{\boldmath $R$}}

\begin{document}

\preprint{}

\title{Key Disclosing in Multiphotons with Quantum Cloning}

\author{Tsuyoshi NISHIOKA}
 \email{nishioka@isl.melco.co.jp}
 \affiliation
 {Information Technology R \& D Center, Mitsubishi Electric 
Corporation\\
5-1-1 Ofuna, Kamakura, Kanagawa 247-8501, JAPAN\\
TEL: +81-467-41-2190 \quad FAX: +81-467-41-2185}
\author{Toshio HASEGAWA}%
 \email{toshio@isl.melco.co.jp}
 \affiliation
 {Information Technology R \& D Center, Mitsubishi Electric 
Corporation\\
5-1-1 Ofuna, Kamakura, Kanagawa 247-8501, JAPAN\\
TEL: +81-467-41-2190 \quad FAX: +81-467-41-2185}

\author{Hirokazu ISHIZUKA}
 \email{ishizuka@isl.melco.co.jp}
 \affiliation
 {Information Technology R \& D Center, Mitsubishi Electric 
Corporation\\
5-1-1 Ofuna, Kamakura, Kanagawa 247-8501, JAPAN\\
TEL: +81-467-41-2190 \quad FAX: +81-467-41-2185}


\begin{abstract}
Multiphoton state in quantum cryptography decreases its security.
Key disclosing with universal quantum cloning machine (UQCM)
is considered in explicit manner. 
Although UQCM cannot make perfect clones, 
there is some invariant quantity between the original photon and
the imperfect clones. The invariant quantity, 
the direction of Stokes parameters, tells us the auxiliary information
leading into key information. The attack, then, corresponds to
some kind of quantum non-demolition measurement.
Its application to recent high-performance
quantum cryptography, Y-00 protocol, is also studied.
\end{abstract}

\maketitle

\section{Introduction}
Quantum cryptography\cite{BB84,GRTZ02} is expected to play an important 
role in near future information security. 
Because its security is based on 
quantum mechanics instead of computational complexity and
is absolutely proved in recent reports\cite{Ma98}-\cite{SP00}. 
Quantum cryptography in real
world has, however, no perfect security, since it consists of 
imperfect devices\cite{Lu00}.
Almost reports\cite{BBBSS92}-\cite{HNIAMT03} on 
their experiments use weak coherent 
pulse and its pulse includes a little multiphoton.
The multiphoton decreases its security. 

In this paper, we study
how to disclose key information in multiphoton using universal
quantum cloning machine(UQCM)\cite{GM97,BH96} in explicit manner.
No-cloning theorem\cite{WZ82}, however, says that
it is impossible to make perfect clones from unknown quantum state
and any UQCM cannot make perfect clones. Imperfect clones do not tell 
us the exact information. Our strategy consists of three steps. The first
step splits multiphoton into two identical (multi-)photons. The second step
amplifies the split one to large number of imperfect clones
with UQCM and measures them for 
`auxiliary' information to be needed in 
the following correct observation. The third step observes
the other split one correctly with the auxiliary information
and gets correct `parity' information which equals to key information
generally. 
The auxiliary information is,
for example, a polarization base in the polarization coding and
corresponds to the direction of Stokes parameters. The parity 
information, then, corresponds to `up' and `down.' Key point of
our attack is that the imperfect clones with low fidelity keep the 
original direction exactly, although their parity information becomes
obscure. Therefore we can measure accurately the direction of the
Stokes parameters with large number of clones.

Recently much higher performance quantum cryptography named 
``Y-00 protocol\cite{Hi03}'' with mesoscopic
coherent pulse is reported\cite{BCKYDPP02}-\cite{HKS02}. Its bit rate
is 1,000-10,000 times higher than that of conventional quantum 
cryptography because its pulse has 100-1,000 photons. The proposers
say that it has enough security by virtue of quantum noise, although
its pulse has multiphoton. 
Its two key ingredients are multi-valued modulation related to 
the auxiliary information and `ciphering wheel' which is a table of
mapping from the parity and the auxiliary information 
to the key information.
The attacker, Eve, knowing the correct parity information, 
gets wrong key information, if  her auxiliary information 
differs from the correct one slightly. 
Disclosing the correct key information needs the accurate auxiliary
information.
We, then, apply our quantum amplification
attack to Y-00 protocol.

The outline of the paper is as follows.
We define qubit and introduce its related quantities in Section II. 
Gisin and Massar's UQCM is introduced in Section III. Quantum 
amplification attack is described in section IV. 
The attack is applied to Y-00 protocol in section V.
Finally we discuss physical background of the quantum amplification 
attack in section VI.

\section{Qubit}
Qubit is generally defined by
\begin{equation}
|\psi\rangle=
\cos\frac{\theta}{2}|H\rangle + \sin\frac{\theta}{2}e^{i\phi}|V\rangle,
\end{equation}
on some basis, where $|H\rangle$ is horizontal linear polarized 
single photon state, $|V\rangle$ is vertical linear polarized 
single photon state, and $0\le\theta\le\pi$, $0\le\phi<2\pi$.
 We focus on photon state as qubit in this paper.

The photon state is represented as a point on Poincar\'{e} sphere
and is parameterized with Stokes parameters. Stokes parameters are
macroscopically defined by
\begin{eqnarray}
S_{1} &=& P_{LH}-P_{LV}, \\
S_{2} &=& P_{L+45}-P_{L-45}, \\
S_{3} &=& P_{RHC}-P_{LHC}, 
\end{eqnarray}
where $P_{LH}$, $P_{LV}$, $P_{L+45}$, $P_{L-45}$, $P_{RHC}$, and 
$P_{LHC}$ are
power of horizontal, vertical, +45 degree, -45 degree linear 
and right-handed, left-handed circular polarized components of light.
These definitions are microscopically equivalent to
\begin{eqnarray}
S_{1} &=& a^{\dagger}_{H}a_{H}-a^{\dagger}_{V}a_{V},\\
S_{2} &=& a_{H}^{\dagger}a_{V}+a_{V}^{\dagger}a_{H},\\
S_{3} &=& i(a_{H}^{\dagger}a_{V}-a_{V}^{\dagger}a_{H}), 
\end{eqnarray}
where $a_{H}$ and $a_{V}$ are annihilation operators of horizontal
and vertical linear polarized mode. These parameters are, then, 
equivalent to angular momentum operators with the following 
commutation relations:
\begin{equation}
[S_{i},S_{j}]=-2i\epsilon_{ijk}S_{k},
\label{CCR}
\end{equation}
where $\epsilon_{ijk}$ is the totally antisymmetric Levi-Civita
symbol and $\epsilon_{123}=1$.
The commutation relations, thus, prevent all Stokes parameters from 
being measured simultaneously and exactly.

The photon state $|\psi\rangle$ has the following density matrix:
\begin{eqnarray}
\rho \!\!&=& \!\!\cos^{2}\frac{\theta}{2}|H\rangle\langle H|
+\cos\frac{\theta}{2}\sin\frac{\theta}{2}e^{-i\phi}|H\rangle\langle V|
\nonumber\\
&& \!\!+\cos\frac{\theta}{2}\sin\frac{\theta}{2}e^{i\phi}|V\rangle\langle H|
+\sin^{2}\frac{\theta}{2}|V\rangle\langle V|.
\end{eqnarray}
The expectation value of Stokes parameters are, then, calculated by
\begin{eqnarray}
\langle S_{1}\rangle &=& {\rm Tr}[S_{1}\rho]=\cos\theta,\\
\langle S_{2}\rangle &=& {\rm Tr}[S_{2}\rho]=\sin\theta\cos\phi,\\
\langle S_{3}\rangle &=& {\rm Tr}[S_{3}\rho]=\sin\theta\sin\phi,
\end{eqnarray}
and their dispersions are given by
\begin{eqnarray}
\langle \Delta S_{1}^{2} \rangle \!\!&=& 
\!\!\langle S_{1}^{2}\rangle-\langle S_{1}\rangle^{2}=1-\cos^{2}\theta,\\
\langle \Delta S_{2}^{2} \rangle \!\!&=& 
\!\!\langle S_{2}^{2}\rangle-\langle S_{2}\rangle^{2}
=1-\sin^{2}\theta\cos^{2}\phi,\\
\langle \Delta S_{2}^{2} \rangle \!\!&=& 
\!\!\langle S_{3}^{2}\rangle-\langle S_{3}\rangle^{2}
=1-\sin^{2}\theta\sin^{2}\phi.
\end{eqnarray}
Measuring Stokes parameters on single photon, therefore, gains no 
sufficient large S/N ratio.

The parity information is defined  by the sign of $S_{2}$ and
the auxiliary information is defined by the quotient space of 
$(\theta,\phi)$ which has the following equivalent class:
\begin{equation}
(\theta,\phi)\equiv(\pi-\theta,\phi+\pi).
\end{equation}

\section{Universal Quantum Cloning Machine}
\begin{figure}[htbp]
\includegraphics
[viewport = 30  180 350 265, clip=true, width=1.40\linewidth]
{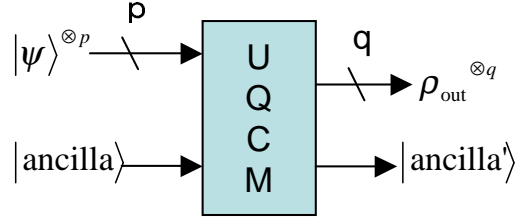}
\caption{\it Gisin and Massar's UQCM}
\label{fig:UQCM}
\end{figure}

\begin{figure}[htbp]
\includegraphics
[viewport = 0  70 330 240, clip=true, width=1.30\linewidth]
{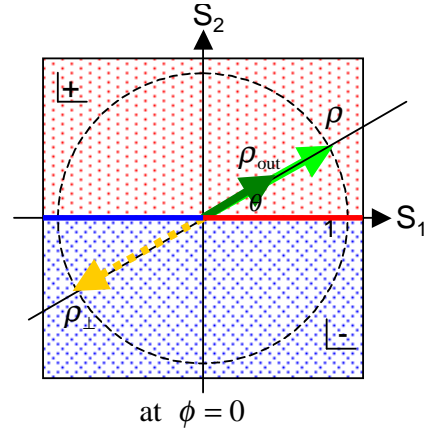}
\caption{\it Clone State}
\label{fig:clone}
\end{figure}

The concept of universal quantum cloning machine (UQCM) is introduced by
Bu\v{z}ek and Hillery\cite{BH96} although No-cloning theorem
prohibits perfect clones. UQCM, then, makes imperfect clones and
`universal' means that the quality of clone does not depend on an input
state. Bu\v{z}ek and Hillery's UQCM making two clones of one qubit
is, however, insufficient for our attack\cite{Ba03} 
and we use Gisin and Massar's
generalized UQCM\cite{GM97} making $q$-identical clones of 
$p$-identical qubits ($q>p$). 
Given the $p$-identical input states $|\psi\rangle$, UQCM 
makes $q$-clones in Fig. \ref{fig:UQCM}. 
\\Their density matrix has the following form:
\begin{equation}
\rho_{out}=F\rho+D\rho_{\bot},
\label{eq:UQCM}
\end{equation}
where $\rho_{\bot}$ is the density matrix of the orthogonal state of 
$|\psi\rangle$ and $F$ is fidelity and $D$ is disturbance.
The fidelity and the disturbance satisfy the relation $F+D=1$.
The orthogonal state with the following form:
\begin{equation}
|\psi_{\bot}\rangle
=\sin\frac{\theta}{2}|H\rangle-\cos\frac{\theta}{2}e^{i\phi}|V\rangle,
\end{equation}
has the diametrical Stokes parameters
\begin{eqnarray}
\langle S_{\bot1}\rangle&=&-\cos\theta,\\
\langle S_{\bot2}\rangle&=&-\sin\theta\cos\phi,\\
\langle S_{\bot3}\rangle&=&-\sin\theta\sin\phi.
\end{eqnarray}
Therefore the direction of Stokes parameters
is invariant in quantum cloning and its length
is shrinking whose factor is given by $\eta=F-D$ in Fig. \ref{fig:clone}.
The UQCM, then, decreases the parity information but conserves 
the auxiliary information.

Gisin and Massar give the following fidelity\cite{GM97}:
\begin{equation}
F=\frac{q(p+1)+p}{q(p+2)}.
\end{equation}
The result includes Bu\v{z}ek and Hillery's result 
at $p=1$ and $q=2$. We,  then, get the following 
disturbance:
\begin{equation}
D=\frac{q-p}{q(p+2)},
\end{equation}
and the shrinking factor is given by
\begin{equation}
\eta=\frac{p(q+2)}{q(p+2)}.
\end{equation}

\section{Quantum Amplification Attack}
\begin{figure}[htbp]
\includegraphics
[viewport = 20  60 450 230, clip=true, width=1.00\linewidth]
{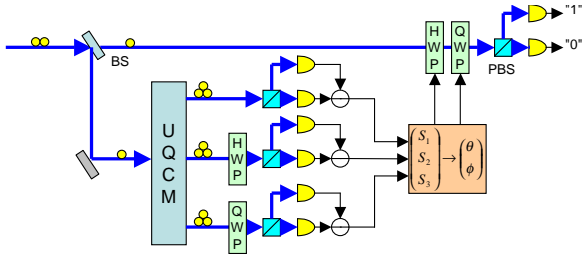}
\caption{\it Quantum Amplification Attack}
\label{fig:QAA}
\end{figure}

We assume the targeted multiphoton can be divided into two 
(multi-) photons.
The attack consists of two phases; the first one is measurement for the 
auxiliary information using the divided (multi-) photon
and the second one is observation for the parity
information using the other divided (multi-) photon. 
The first phase is divided into two steps; Step 1 is
quantum amplification of the former (multi-) photon and Step 2 is
measurement of the quantum amplified
photons for the auxiliary information. 
The whole attack, then, consists of three steps in 
Fig. \ref{fig:QAA}.

\subsection{Quantum amplification}
Eve makes Gisin and Massar's UQCM work in $L$-steps cascade 
manner and propagates the imperfect clones to $(q/p)^{L}$-photons. 
Assuming that the density matrix in the  $k$-step is given by
\begin{equation}
\rho_{k}= a_{k}\rho+b_{k}\rho_{\bot},
\end{equation} 
the coefficients $a_{k}$ and $b_{k}$ obey the following recurrences:
\begin{equation}
a_{k+1}=Fa_{k}+Db_{k},
\end{equation}
\begin{equation}
b_{k+1}=Da_{k}+Fb_{k},
\end{equation}
from (\ref{eq:UQCM}).
Solving the recurrences, we get the following final state:
\begin{equation}
\rho_{\rm final}=\frac{1}{2}\left(1+\eta^{L}\right)\rho
+\frac{1}{2}\left(1-\eta^{L}\right)\rho_{\bot}.
\end{equation}
The final state is much depolarized and its parity information
cannot be gained, if $L$ is large enough .

\subsection{Measurement for the auxiliary information}
Eve measures Stokes parameters of the $(q/p)^{L}$-clones 
though the states are much depolarized.
The expectation values of Stokes parameters per single photon are
calculated by
\begin{eqnarray}
\langle S_{1}\rangle &=& 
{\rm Tr}[S_{1}\rho_{\rm final}]=\eta^{L}\cos\theta,\\
\langle S_{2}\rangle &=& 
{\rm Tr}[S_{2}\rho_{\rm final}]=\eta^{L}\sin\theta\cos\phi,\\
\langle S_{3}\rangle &=& 
{\rm Tr}[S_{3}\rho_{\rm final}]=\eta^{L}\sin\theta\sin\phi,
\end{eqnarray}
and their dispersions are given by
\begin{eqnarray}
\Delta S_{1}^{2} &=& 1-\eta^{2L}\cos^{2}\theta,\\
\Delta S_{2}^{2} &=& 
1-\eta^{2L}\sin^{2}\theta\cos^{2}\phi,\\
\Delta S_{3}^{2} &=& 
1-\eta^{2L}\sin^{2}\theta\sin^{2}\phi.
\end{eqnarray}
The whole Stokes parameters 
and the whole dispersions are proportional to 
the total photon number $(q/p)^{L}$. The whole Stokes
parameters are obtained by
\begin{eqnarray}
S_{1}^{\rm total} &=& 
\left(\frac{q}{p}\right)^{L}\!\!\!\langle S_{1}\rangle
=\left(\frac{q+2}{p+2}\right)^{L}\cos\theta,\\
S_{2}^{\rm total} &=& 
\left(\frac{q+2}{p+2}\right)^{L}\sin\theta\cos\phi,\\
S_{3}^{\rm total} &=& 
\left(\frac{q+2}{p+2}\right)^{L}\sin\theta\sin\phi.
\end{eqnarray}
Their S/N ratio is also estimated by
\begin{equation}
S/N
=\frac{(q/p)^{L}\langle S_{\ast}\rangle}{(q/p)^{L/2}\Delta S_{\ast}}
\approx \left(\frac{q}{p}\right)^{\!\!\!\frac{L}{2}}\!\!\!\eta^{L}
=\left(\frac{p(q+2)^{2}}{q(p+2)^{2}}\right)^{\!\!\!\frac{L}{2}},
\label{eq:SNQ}
\end{equation}
and then the S/N ratio grows large if the index satisfies the inequality
\begin{equation}
\frac{p(q+2)^{2}}{q(p+2)^{2}}>1.
\end{equation}
The growing condition is as follows:
\begin{equation}
q+\frac{4}{q}>p+\frac{4}{p},
\end{equation}
and its solution becomes
\begin{equation}
q>4\;{\rm at}\;p=1,\quad {\rm or}\quad q>p\ge2.
\label{eq:growing}
\end{equation}
 Therefore the whole Stokes parameters 
grow large enough to be measured accurately 
if $L$ is sufficiently large and the inequality (\ref{eq:growing})
is satisfied. Bu\v{z}ek and Hillery's UQCM does not satisfy the
condition and cannot be used in the attack\cite{Ba03}.
Eve, then, measures the Stokes parameters  
and gets the accurate auxiliary information.

We can also estimate statistical S/N ratio although the S/N ratio by quantum
noise has been estimated in the above argument. The final states 
 are $(q/p)^{L}$-identical states obeying binominal distribution
\begin{equation}
P(k)=_{N}\!\!C_{k}
F_{\rm final}\;^{N-k}
D_{\rm final}\;^{k},
\end{equation}
where $P(k)$ is probability distribution function with 
the $N-k$-original state $\rho$ and the $k$-orthogonal state 
$\rho_{\bot}$, $N=(q/p)^{L}$, $F_{\rm final}=(1+\eta^{L})/2$, 
$D_{\rm final}=(1-\eta^{L})/2$, and $_{N}C_{k}$ is binominal
coefficient.  The mean value of Stokes parameter is, then,  
calculated by
\begin{eqnarray}
\langle S_{1}\rangle_{\rm statistics} &=& 
\sum_{k=0}^{N}((N-k)-k)P(k)\cos\theta,\nonumber \\
&=& (1-2D_{\rm final})N\cos\theta,\nonumber\\
&=& \left(\frac{q}{p}\right)^{L}\eta^{L}\cos\theta.
\end{eqnarray}
Its variance is given by
\begin{eqnarray}
\langle \Delta S_{1}^{2}\rangle_{\rm statistics}
&=& \langle S_{1}^{2}\rangle_{\rm statistics}
    - \langle S_{1}\rangle_{\rm statistics}^{2},\nonumber\\
&=& 4F_{\rm final}D_{\rm final}N\cos^{2}\theta.
\end{eqnarray}
Therefore the statistical S/N ratio is obtained by
\begin{eqnarray}
S/N_{\rm statistics}&=&
\frac{\eta^{L}N}{\sqrt{4F_{\rm final}D_{\rm final}N}},\nonumber\\
&=&\frac{\eta^{L}}{\sqrt{1-\eta^{2L}}}\left(\frac{q}{p}\right)^{L}
,\nonumber\\
&\sim& \left( \frac{p(q+2)^{2}}{q(p+2)^{2}} \right)^{\frac{L}{2}}.
\end{eqnarray}
The result is the same as (\ref{eq:SNQ}) in the quantum noise.

\subsection{Observation for the parity information}
Eve observes the latter (multi-)photon with the auxiliary information
obtained in step 2 and gets the correct parity information.
The parity information is equivalent to key information generally 
and she obtains the correct key information.

The attack enables Eve to disclose key information before the auxiliary
information is opened by the legitimate entities and is effective
for some protocols with multiphoton
having no public announcement of the auxiliary
information\cite{HKH98}.

\section{Application to Y-00 Protocol}
Recently reported Y-00 protocol\cite{Hi03}-\cite{HKS02} is 
a kind of quantum key expansion protocol and
1,000-10,000 times faster than conventional quantum cryptography 
because it uses mesoscopic coherent pulse including 100-1,000 
photons. The quantum amplification attack seems to be 
applicable to Y-00 protocol 
whose security depends on quantum noise with two important 
ingredients: multi-valued modulation and ciphering wheel.

\subsection{Essence of Y-00 protocol}
We roughly sketch the essence of Y-00 protocol with polarization 
coding in the following.
Y-00 protocol prepares the $M$-pair states:
\begin{eqnarray}
|+,k\rangle\!\!\!\!&=&\!\!\!\!
\left|\alpha\cos\frac{\theta_{k}}{2}\right\rangle_{H}
\left|\alpha\sin\frac{\theta_{k}}{2}\right\rangle_{V},\\
|-,k\rangle\!\!\!\!&=&\!\!\!\!
\left|\alpha\cos\frac{\theta_{k}+\pi}{2}\right\rangle_{H}
\left|\alpha\sin\frac{\theta_{k}+\pi}{2}\right\rangle_{V},
\end{eqnarray}
where $k=0,\dots,M-1$ and the right-sided ket vector is
coherent state on each basis and $|\alpha|^{2}$ equals to the 
average photon number and $\theta_{k}=\pi k/M$. 
The two pairing states have the diametrical relation with each other
on Poincar\'{e} sphere. The sign in the left side corresponds
to the parity information and the multi-valued $k$ in 
the left side corresponds 
to the auxiliary information. 

Key information is not the same as the parity
information in Y-00 protocol. The ciphering wheel maps from the parity
information and the auxiliary information to the key information in the
following:
\begin{eqnarray}
CW(+,k:{\rm even})&=& 0,\\
CW(-,k:{\rm even})&=& 1,\\
CW(+,k:{\rm odd})&=& 1,\\
CW(-,k:{\rm odd})&=& 0.
\end{eqnarray}
Eve without knowing the auxiliary information, 
thus, cannot guess the correct key
information even though she knows the correct parity information.

The state $|+,k\rangle$ has the following Stokes parameters:
\begin{eqnarray}
\langle S_{1}\rangle &=& |\alpha|^{2}\cos\theta,\\
\langle S_{2}\rangle &=& |\alpha|^{2}\sin\theta,\\
\langle S_{3}\rangle &=& 0,
\end{eqnarray}
and their dispersions are given by
\begin{equation}
\Delta S_{1}^{2}=\Delta S_{2}^{2}=|\alpha|^{2}.
\end{equation}
The neighboring states with the 
same parity cannot be discriminated 
if $M$ is sufficiently large, where the neighboring condition is
given by
\begin{equation}
|\vec{S}_{\pm, k}-\vec{S}_{\pm,k'}|<\Delta S=|\alpha|.
\end{equation}
The condition is equivalent to
\begin{equation}
\Delta k < \frac{M}{\pi|\alpha|},
\end{equation}
and Y-00 protocol seems to be secure if $M > |\alpha|\pi$ is satisfied.

\subsection{Quantum Amplification Attack to Y-00 protocol}
The quantum amplification attack needs no
auxiliary information opened publicly and then seems to be effective
against Y-00 protocol. It uses mesoscopic coherent pulse instead of
single photon state. Eve must extract some single photon states 
from the mesoscopic coherent state in order to let the UQCM work.

She splits the targeted state into two mesoscopic coherent state by
 a beam splitter at first. The one state 
$|\alpha_{1}\cos\theta/2\rangle_{H}|\alpha_{1}\sin\theta/2\rangle_{V}$
is used in the quantum amplification and measurement for the auxiliary
information. 

The other state
$|\alpha_{2}\cos\theta/2\rangle_{H}|\alpha_{2}\sin\theta/2\rangle_{V}$
is used in measurement for the parity information.

Eve, moreover, splits the first state into $J$-weak 
coherent states by beam splitters
in a cascade way
\begin{equation}
\left[
\left|\frac{\alpha_{1}}{\sqrt{J}}\cos\frac{\theta}{2}\right\rangle_{H}
\left|\frac{\alpha_{1}}{\sqrt{J}}\sin\frac{\theta}{2}\right\rangle_{V}
\right]^{\otimes J}
,
\end{equation}
where $|\alpha_{1}|^{2}/J\ll 1$.

One of the weak coherent states is equivalent to
\begin{eqnarray}
|0\rangle_{H}|0\rangle_{V}
+\frac{\alpha}{\sqrt{J}}\left(\cos\frac{\theta}{2}|1\rangle_{H}|0\rangle_{V}
+\sin\frac{\theta}{2}|0\rangle_{H}|1\rangle_{V}\right)&&\nonumber\\
+{\mathcal O}(\frac{|\alpha|^{2}}{J}),
\end{eqnarray}
where $|0\rangle,|1\rangle$ are numbering states in each mode. The first term
is a vacuum and the second is single photon state. Other multiphoton states
are negligible.

The single photon state can be represented by
\begin{equation}
\frac{\alpha_{1}}{\sqrt{J}}\left(\cos\frac{\theta}{2}|H\rangle
+\sin\frac{\theta}{2}|V\rangle\right),
\end{equation}
in qubit-like representation. 
Eve expects to get 
$|\alpha_{1}|^{2}$-identical single photons
because she has $J$-states.
The state, moreover, keeps the auxiliary information $\theta$ perfectly.

Eve quantum-amplifies the obtained single photon states and
measures them for the accurate auxiliary information.
She obtains the correct parity information using the
auxiliary information finally.

\section{Discussion}
In the quantum amplification attack,  the UQCM plays an important role
though any UQCM cannot make perfect clones. Because the UQCM, which is
some kind of unitary transformation, has invariant subspace
in the space of
Stokes parameters $\bR^{3}$.
Although the fidelity of an imperfect 
clone to the original photon decreases below 1, the direction of its
Stokes parameters is invariant between the original and the clone.
The space of Stokes parameters is,  then, divided by one-dimensional
real space and two-dimensional projective space,
\begin{equation}
\bR^{3}\sim \bR\times P^{2},
\end{equation}
where the parity information corresponds to 
the sign of the one-dimensional
real space $\bR$ and the auxiliary information corresponds to the
two-dimensional projective space $P^{2}$
which is invariant under the transformation
by the UQCM.

\begin{figure}[htbp]
\includegraphics
[viewport = 40  70 440 200, clip=true, width=1.00\linewidth]
{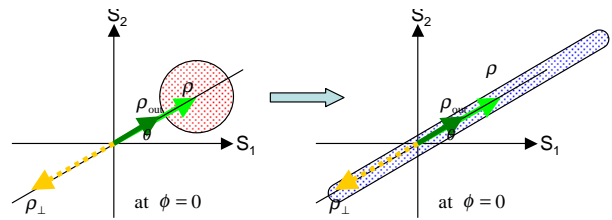}
\caption{\it Squeezing}
\label{fig:squeeze}
\end{figure}

The commutation relations (\ref{CCR}) among Stokes parameters
say that the all parameters cannot be observed simultaneously and
accurately by uncertainty relations.
 It, then, seems strange that the attack measures the accurate
direction of the Stokes parameters. The accurate measurement, 
however, does not contradict the uncertainty relations
because the attack gains no parity information and then two diametrical
points on Poincar\'{e} sphere cannot be discriminated.
The attack, then, squeezes any state to extending to the parity real space
and to shrinking in the projective space in Fig. \ref{fig:squeeze}. 
Therefore the attack may be 
regarded as some kind of quantum non-demolition measurement\cite{BK96}.

The attack is applicable to BB84 protocol with multiphoton although
it does not work in the protocol with single photon because
the protocol carries critical information on the parity information. 
On the other hand, the attack
is also expected to be effective in Y-00 protocol because
the protocol carries critical information on the auxiliary information
rather than the parity information.

The attack has, however, some open problems. It is generally said that
no photon-number amplifier can avoid fluctuation of photon-number
and then sufficient large S/N ratio cannot be gained\cite{Ba03}.
The UQCM is, however, a kind of unitary transformation and 
the photon-number is conserved. It, then, seems to be hard for the UQCM
to exist. It is more severe problem that
the measurement of Strokes parameters is 
executed to the whole clones and the treatment as the mixed states
is somewhat wondered since 
the output state of the UQCM is entangled state and independent 
measurement changes the entangled output into the mixed states.
 The cascade manner operation of the UQCM
is not a severe problem because the attack is effective at $L=1$.
Its application to B92 protocol\cite{Be92} whose critical information is  
also carried on the auxiliary information would be challenging
because recent reports\cite{QC02,TKI02} have proved its unconditional
security.
\begin{acknowledgments}
We thank Prof. Barbosa for helpful discussion though he doubts that
there is such a UQCM that conserves photon-number without 
photon-number fluctuation.
This work was supported by the project on ``Research and Development
on Quantum Cryptography'' of Telecommunications Advancement 
Organization as part of the programme ``Research and Development on
Quantum Communication Technology'' of the Ministry of Public 
Management, Home Affairs, Posts and Telecommunications of Japan.
\end{acknowledgments}


\end{document}